\documentstyle[12pt,epsf]{article}
\epsfverbosetrue
\def\figlabel#1{\xdef#1{\thefigure}}

\def\figalign#1#2#3#4#5#6{
\begin{figure}
\centerline{
\hbox to 2.5truein{\vtop{\hsize=2.5truein\epsfxsize=6cm
\centerline{\epsfbox{#1} }
\caption[]{#3}
\figlabel{#2} }}
\qquad\hbox to 2.5truein{\vtop{\hsize=2.5truein\epsfxsize=6cm
\centerline{\epsfbox{#4} }
\caption[]{#6}
\figlabel{#5} }} }
\end{figure} }

\newcommand{\beq}{\begin{equation}}
\newcommand{\eeq}{\end{equation}}
\newcommand{\bear}{\begin{eqnarray}}
\newcommand{\eear}{\end{eqnarray}}

\def\tr{{\rm Tr}}

\newcommand{\k}{{\cal K}}
\newcommand{\ie}{{\it i.e.}}

\linethickness{.05 pt}
\newsavebox{\unoj}
\savebox{\unoj}(10,3){
\put(2,5) {\circle{10}}
\put(2,0){\line(0,0){10}
\put(0,4){${\scriptscriptstyle j}$}} }
\newcommand{\cunoj}{\mbox{\usebox{\unoj} \hskip 15pt  } } 

\newsavebox{\unodos}
\savebox{\unodos}(10,3){
\put(2,5) {\circle{10}}
\put(2,0){\line(0,0){10}
\put(0,4){${\scriptscriptstyle 2}$}} }
\newcommand{\cunodos}{\mbox{\usebox{\unodos} \hskip 15pt  } } 

\newsavebox{\unotres}
\savebox{\unotres}(10,3){
\put(2,5) {\circle{10}}
\put(2,0){\line(0,0){10}
\put(0,4){${\scriptscriptstyle 3}$}} }
\newcommand{\cunotres}{\mbox{\usebox{\unotres} \hskip 15pt  } } 

\newsavebox{\unocuatro}
\savebox{\unocuatro}(10,3){
\put(2,5) {\circle{10}}
\put(2,0){\line(0,0){10}
\put(0,4){${\scriptscriptstyle 4}$}} }
\newcommand{\cunocuatro}{\mbox{\usebox{\unocuatro} \hskip 15pt  } } 

\newsavebox{\dosi}
\savebox{\dosi}(10,3){
\put(2,5) {\circle{10}}
\put(0,0.3){\line(0,1){9.2}}
\put(4,0.3){\line(0,1){9.2} } }

\newsavebox{\dosii}
\savebox{\dosii}(10,3){
\put(2,5) {\circle{10}}
\put(-3,5){\line(3,0){10}}
\put(2,0){\line(0,5){10} } }
\newcommand{\cdosii}{\mbox{\usebox{\dosii} \hskip 15pt  } } 

\newsavebox{\dosiidosuno}
\savebox{\dosiidosuno}(10,3){
\put(2,5) {\circle{10}}
\put(-3,5){\line(3,0){10}}
\put(2,0){\line(0,5){10}
\put(0,7){${\scriptscriptstyle 2}$} } }
\newcommand{\cdosiidosuno}{\mbox{\usebox{\dosiidosuno} \hskip 15pt  } } 

\newsavebox{\dosiidosdos}
\savebox{\dosiidosdos}(10,3){
\put(2,5) {\circle{10}}
\put(-3,5){\line(3,0){10}}
\put(2,0){\line(0,5){10}
\put(0,7){${\scriptscriptstyle 2}$}
\put(-3,5){${\scriptscriptstyle 2}$} } }
\newcommand{\cdosiidosdos}{\mbox{\usebox{\dosiidosdos} \hskip 15pt  } } 

\newsavebox{\dosiitresuno}
\savebox{\dosiitresuno}(10,3){
\put(2,5) {\circle{10}}
\put(-3,5){\line(3,0){10}}
\put(2,0){\line(0,5){10}
\put(0,7){${\scriptscriptstyle 3}$} } }
\newcommand{\cdosiitresuno}{\mbox{\usebox{\dosiitresuno} \hskip 15pt  } } 

\newsavebox{\tresiii}
\savebox{\tresiii}(10,3){
\put(2,5) {\circle{10}}
\put(0,0.3){\line(0,1){9.2}}
\put(-3,5){\line(3,0){10}}
\put(4,0.4){\line(0,1){9.1} } }
\newcommand{\ctresiii}{\mbox{\usebox{\tresiii} \hskip 15pt  }
} 

\newsavebox{\tresiv}
\savebox{\tresiv}(10,3){
\put(2,5) {\circle{10}}
\put(-0.25,0.5){\line(1,2){4.45}}
\put(-3,5){\line(3,0){10}}
\put(4.25,0.6){\line(-1,2){4.45} } }
\newcommand{\ctresiv}{\mbox{\usebox{\tresiv} \hskip 15pt  }
} 

\newsavebox{\tresivdos}
\savebox{\tresivdos}(10,3){
\put(2,5) {\circle{10}}
\put(-0.25,0.5){\line(1,2){4.45}}
\put(-3,5){\line(3,0){10}}
\put(4.25,0.6){\line(-1,2){4.45}
\put(-4,4.5){${\scriptscriptstyle 2}$} } }
\newcommand{\ctresivdos}{\mbox{\usebox{\tresivdos} \hskip 15pt  }
} 

\newsavebox{\tresiiidos}
\savebox{\tresiiidos}(10,3){
\put(2,5) {\circle{10}}
\put(0,0.3){\line(0,1){9.2}}
\put(-3,5){\line(3,0){10}}
\put(4,0.4){\line(0,1){9.1}
\put(-4,6){${\scriptscriptstyle 2}$} } }
\newcommand{\ctresiiidos}{\mbox{\usebox{\tresiiidos} \hskip 15pt  }
} 

\newsavebox{\tresiiidosbis}
\savebox{\tresiiidosbis}(10,3){
\put(2,5) {\circle{10}}
\put(0,0.3){\line(0,1){9.2}}
\put(-3,5){\line(3,0){10}}
\put(4,0.4){\line(0,1){9.1}
\put(-2.5,5){${\scriptscriptstyle 2}$} } }
\newcommand{\ctresiiidosbis}{\mbox{\usebox{\tresiiidosbis} \hskip 15pt  }
} 

\newsavebox{\cuavi}
\savebox{\cuavi}(10,3){
\put(2,5) {\circle{10}}
\put(-1,0.8){\line(0,1){8.2}}
\put(2,0){\line(0,5){10} }
\put(5,1){\line(0,1){7.8} }
\put(-3,5){\line(3,0){10} }}
\newcommand{\ccuavi}{\mbox{\usebox{\cuavi} \hskip 15pt  } } 

\newsavebox{\cuavii}
\savebox{\cuavii}(10,3){
\put(2,5) {\circle{10}}
\put(-1,0.8){\line(0,1){8.1}}
\put(-2.7,3.1){\line(5,6){5.6} }
\put(6.5,3.1){\line(-5,6){5.6} }
\put(5,1){\line(0,1){7.8} } }
\newcommand{\ccuavii}{\mbox{\usebox{\cuavii} \hskip 15pt  }
} 

\newsavebox{\cuaviii}
\savebox{\cuaviii}(10,3){
\put(2,5) {\circle{10}}
\put(-1,0.8){\line(0,1){8.2}}
\put(2,0){\line(1,3){2.99} }
\put(5,1){\line(-1,3){3} }
\put(-3,5){\line(3,0){10} }}
\newcommand{\ccuaviii}{\mbox{\usebox{\cuaviii} \hskip 15pt  }
}

\newsavebox{\cuaix}
\savebox{\cuaix}(10,3){
\put(2,5) {\circle{10}}
\put(0,0.3){\line(0,1){9.2}}
\put(4,0.3){\line(0,1){9.2} }
\put(-2.7,3){\line(3,0){9.2}}
\put(-2.7,7){\line(3,0){9.2}} }
\newcommand{\ccuaix}{\mbox{\usebox{\cuaix} \hskip 15pt  }
} 

\newsavebox{\cuax}
\savebox{\cuax}(10,3){
\put(2,5) {\circle{10}}
\put(-0.3,0.5){\line(1,2){4.5}}
\put(4.3,0.6){\line(-1,2){4.45} }
\put(-2.7,3){\line(3,0){9.2}}
\put(-2.7,7){\line(3,0){9.2}} }
\newcommand{\ccuax}{\mbox{\usebox{\cuax} \hskip 15pt  }
} 

\newsavebox{\cuaxi}
\savebox{\cuaxi}(10,3){
\put(2,5) {\circle{10}}
\put(-3,5){\line(3,0){10}}
\put(2,0){\line(0,5){10} }
\put(-1.6,1.4){\line(1,1){7.1} }
\put(5.5,1.5){\line(-1,1){7.05} }}
\newcommand{\ccuaxi}{\mbox{\usebox{\cuaxi} \hskip 15pt  } }

\linethickness{.05 pt}

\newsavebox{\diba}
\newsavebox{\dibb}
\newsavebox{\dibc}
\newsavebox{\dibd}
\newsavebox{\dibe}
\newsavebox{\dibf}
\newsavebox{\dibg}
\newsavebox{\dibh}
\newsavebox{\dibi}
\newsavebox{\dibj}
\newsavebox{\dibk}
\newsavebox{\dibl}
\newsavebox{\dibm}
\newsavebox{\dibn}
\newsavebox{\dibo}
\newsavebox{\dibp}
\newsavebox{\dibq}
\newsavebox{\dibr}
\newsavebox{\dibs}
\newsavebox{\dibt}
\newsavebox{\dibu}
\newsavebox{\dibv}
\newsavebox{\dibw}
\newsavebox{\dibx}
\newsavebox{\diby}
\newsavebox{\dibz}

\savebox{\diba}(0,2){
\put(2,2) {\circle{4}}
\put(2,0){\line(0,1){4}} }

\savebox{\dibb}(0,2){
\put(2,2) {\circle{4}}
\put(1.5,0.1){\line(0,1){3.9}}
\put(2.5,0.1){\line(0,1){3.9}} }

\savebox{\dibc}(0,2){
\put(2,2) {\circle{4}}
\put(2,0){\line(0,1){4}} 
\put(0,2){\line(1,0){4}} }

\savebox{\dibd}(0,2){
\put(2,2) {\circle{4}}
\put(.2,1){\line(1,0){3.45}}
\put(0.1,1.7){\line(2,1){3.5}}
\put(.6,3.4){\line(2,-1){3.5}}  }

\savebox{\dibe}(0,2){
\put(2,2) {\circle{4}}
\put(2,0){\line(0,1){4}}
\put(1,0.2){\line(0,1){3.5}}
\put(3,0.2){\line(0,1){3.5}}  }

\savebox{\dibf}(0,2){
\put(2,2) {\circle{4}}
\put(0 ,2){\line(1,0){4}}
\put(1,0.2){\line(0,1){3.5}}
\put(3,0.2){\line(0,1){3.5}}  }

\savebox{\dibg}(0,2){
\put(2,2) {\circle{4}}
\put(2,0){\line(0,1){4}}
\put(0.2,1.2){\line(2,1){3.6}}
\put(.2,2.9){\line(2,-1){3.6}} }

\savebox{\dibh}(0,0){
\put(2,2) {\circle{4}}
\put(0,2){\line(1,0){4}}
\put(2,0){\line(0,1){2.0}} }

\savebox{\dibi}(0,2){
\put(2,2) {\circle{4}}
\put(0,2){\line(1,0){4}}
\put(2,0){\line(0,1){2.0}}
\put(.6,3.4){\line(1,0){2.8}} }

\savebox{\dibj}(0,0){
\put(2,2) {\circle{4}}
\put(0 ,2){\line(1,0){4}}
\put(1,.25){\line(0,1){1.8}}
\put(3,.25){\line(0,1){1.8}}  }

\savebox{\dibk}(0,2){
\put(2,2) {\circle{4}}
\put(.5,.65){\line(0,1){2.7}}
\put(1.5,.2){\line(0,1){3.8}}
\put(2.5,.2){\line(0,1){3.8}}
\put(3.5,.65){\line(0,1){2.7}} }

\savebox{\dibl}(0,2){
\put(2,2) {\circle{4}}
\put(.6,.6){\line(1,0){2.75}}
\put(.25,1){\line(1,0){3.5}}
\put(0.1,1.7){\line(2,1){3.5}}
\put(.55,3.35){\line(2,-1){3.5}} }

\savebox{\dibm}(0,2){
\put(2,2) {\circle{4}}
\put(0,2){\line(1,0){4}}
\put(0.1,1.5){\line(1,0){3.9}}
\put(.25,.9){\line(2,1){3.7}}
\put(.65,3.4){\line(1,0){2.75}} }

\savebox{\dibn}(0,2){
\put(2,2) {\circle{4}}
\put(0.1,2.2){\line(3,1){3.5}}
\put(.55,3.35){\line(3,-1){3.5}}
\put(0.6,.6){\line(3,1){3.5}}
\put(.15,1.75){\line(3,-1){3.5}} }

\savebox{\dibo}(0,2){
\put(2,2) {\circle{4}}
\put(2,0){\line(0,1){4}}
\put(1,0.2){\line(0,1){3.5}}
\put(3,0.2){\line(0,1){3.5}}
\put(0 ,2){\line(1,0){4}}  }

\savebox{\dibp}(0,2){
\put(2,2) {\circle{4}}
\put(.45,.7){\line(0,1){2.6}}
\put(0,2){\line(1,0){4}}
\put(.65,.6){\line(1,0){2.75}}
\put(2,0){\line(0,1){4}} }

\savebox{\dibq}(0,2){
\put(2,2) {\circle{4}}
\put(.6,.65){\line(1,0){2.8}}
\put(2,0){\line(0,1){4}}
\put(0.1,1.7){\line(2,1){3.5}}
\put(.55,3.35){\line(2,-1){3.5}} }

\savebox{\dibr}(0,2){
\put(2,2) {\circle{4}}
\put(1.5,.1){\line(0,1){3.9}}
\put(2.5,.1){\line(0,1){3.9}}
\put(.1,1.5){\line(1,0){3.9}}
\put(.1,2.5){\line(1,0){3.9}} }

\savebox{\dibs}(0,2){
\put(2,2) {\circle{4}}
\put(1.5,.1){\line(0,1){3.9}}
\put(2.5,.1){\line(0,1){3.9}}
\put(0.25,1.1){\line(2,1){3.5}}
\put(.25,2.9){\line(2,-1){3.5}} }

\savebox{\dibt}(0,2){
\put(2,2) {\circle{4}}
\put(0.25,1.1){\line(2,1){3.5}}
\put(.25,2.9){\line(2,-1){3.5}}
\put(0,2){\line(1,0){4}}
\put(2,0){\line(0,1){4}} }

\savebox{\dibu}(0,0){
\put(2,2) {\circle{4}}
\put(.2,1.5){\line(1,0){3.8}}
\put(2,0){\line(0,1){1.5}}
\put(.2,2.2){\line(1,0){3.8}}
\put(.4,3){\line(1,0){3.2}}}

\savebox{\dibv}(0,0){
\put(2,2) {\circle{4}}
\put(.2,1.5){\line(1,0){3.8}}
\put(2,0){\line(0,1){1.5}}
\put(.2,2.2){\line(1,0){3.8}}
\put(2,2.2){\line(0,1){1.8}}}

\savebox{\dibw}(0,0){
\put(2,2) {\circle{4}}
\put(0,2){\line(1,0){4}}
\put(2,0){\line(0,1){2.0}}
\put(1,.3){\line(0,1){1.7}}
\put(3,.25){\line(0,1){1.7}} }

\savebox{\dibx}(0,0){
\put(2,2) {\circle{4}}
\put(.3,1.2){\line(1,0){3.6}}
\put(1.2,1.2){\line(0,1){1.6}}
\put(.3,2.8){\line(1,0){3.6}}
\put(2.8,1.2){\line(0,1){1.6}}}

\savebox{\diby}(0,0){
\put(2,2) {\circle{4}}
\put(0 ,2){\line(1,0){4}}
\put(1,.25){\line(0,1){1.8}}
\put(3,.25){\line(0,1){1.8}}
\put(.3,3){\line(1,0){3.4}} }

\savebox{\dibz}(0,2){
\put(2,2) {\circle{4}}
\put(0,2){\line(1,0){4}}
\put(.6,3.4){\line(1,0){2.7}}
\put(0.25,1.1){\line(2,1){3.5}}
\put(.25,2.9){\line(2,-1){3.5}} }

\newsavebox{\faca}
\newsavebox{\facy}
\newsavebox{\facb}
\newsavebox{\facc}
\newsavebox{\facd}
\newsavebox{\face}
\newsavebox{\facf}
\newsavebox{\facg}
\newsavebox{\fach}
\newsavebox{\faci}
\newsavebox{\facj}
\newsavebox{\fack}
\newsavebox{\facl}
\newsavebox{\facm}
\newsavebox{\facn}
\newsavebox{\faco}
\newsavebox{\facp}
\newsavebox{\facq}
\newsavebox{\facr}
\newsavebox{\facs}
\newsavebox{\fact}
\newsavebox{\facu}
\newsavebox{\facv}
\newsavebox{\facw}
\newsavebox{\facx}

\savebox{\faca}(0,2){
\put(2,2) {\circle{4}}
\put(2,0){\line(0,1){4}} }

\savebox{\facb}(0,2){
\put(2,2) {\circle{4}}
\put(1.5,0.1){\line(0,1){3.9}}
\put(2.5,0.1){\line(0,1){3.9}} }

\savebox{\facc}(0,2){
\put(2,2) {\circle{4}}
\put(2,0){\line(0,1){4}} 
\put(0,2){\line(1,0){4}} }

\savebox{\facd}(0,2){
\put(2,2) {\circle{4}}
\put(.2,1){\line(1,0){3.45}}
\put(0.1,1.7){\line(2,1){3.5}}
\put(.6,3.4){\line(2,-1){3.5}}  }

\savebox{\face}(0,2){
\put(2,2) {\circle{4}}
\put(2,0){\line(0,1){4}}
\put(1,0.2){\line(0,1){3.5}}
\put(3,0.2){\line(0,1){3.5}}  }

\savebox{\facf}(0,2){
\put(2,2) {\circle{4}}
\put(0 ,2){\line(1,0){4}}
\put(1,0.2){\line(0,1){3.5}}
\put(3,0.2){\line(0,1){3.5}}  }

\savebox{\facg}(0,2){
\put(2,2) {\circle{4}}
\put(2,0){\line(0,1){4}}
\put(0.2,1.2){\line(2,1){3.6}}
\put(.2,2.9){\line(2,-1){3.6}} }

\savebox{\fach}(0,0){
\put(2,2) {\circle{4}}
\put(0,2){\line(1,0){4}}
\put(2,0){\line(0,1){2.0}} }

\savebox{\faci}(0,2){
\put(2,2) {\circle{4}}
\put(0,2){\line(1,0){4}}
\put(2,0){\line(0,1){2.0}}
\put(.6,3.4){\line(1,0){2.8}} }

\savebox{\facj}(0,0){
\put(2,2) {\circle{4}}
\put(0 ,2){\line(1,0){4}}
\put(1,.25){\line(0,1){1.8}}
\put(3,.25){\line(0,1){1.8}}  }

\savebox{\fack}(0,2){
\put(2,2) {\circle{4}}
\put(.5,.65){\line(0,1){2.7}}
\put(1.5,.2){\line(0,1){3.8}}
\put(2.5,.2){\line(0,1){3.8}}
\put(3.5,.65){\line(0,1){2.7}} }

\savebox{\facl}(0,2){
\put(2,2) {\circle{4}}
\put(.6,.6){\line(1,0){2.75}}
\put(.25,1){\line(1,0){3.5}}
\put(0.1,1.7){\line(2,1){3.5}}
\put(.55,3.35){\line(2,-1){3.5}} }

\savebox{\facm}(0,2){
\put(2,2) {\circle{4}}
\put(0,2){\line(1,0){4}}
\put(0.1,1.5){\line(1,0){3.9}}
\put(.25,.9){\line(2,1){3.7}}
\put(.65,3.4){\line(1,0){2.75}} }

\savebox{\facn}(0,2){
\put(2,2) {\circle{4}}
\put(0.1,2.2){\line(3,1){3.5}}
\put(.55,3.35){\line(3,-1){3.5}}
\put(0.6,.6){\line(3,1){3.5}}
\put(.15,1.75){\line(3,-1){3.5}} }

\savebox{\faco}(0,2){
\put(2,2) {\circle{4}}
\put(2,0){\line(0,1){4}}
\put(1,0.2){\line(0,1){3.5}}
\put(3,0.2){\line(0,1){3.5}}
\put(0 ,2){\line(1,0){4}}  }

\savebox{\facp}(0,2){
\put(2,2) {\circle{4}}
\put(.45,.7){\line(0,1){2.6}}
\put(0,2){\line(1,0){4}}
\put(.65,.6){\line(1,0){2.75}}
\put(2,0){\line(0,1){4}} }

\savebox{\facq}(0,2){
\put(2,2) {\circle{4}}
\put(.6,.65){\line(1,0){2.8}}
\put(2,0){\line(0,1){4}}
\put(0.1,1.7){\line(2,1){3.5}}
\put(.55,3.35){\line(2,-1){3.5}} }

\savebox{\facr}(0,2){
\put(2,2) {\circle{4}}
\put(1.5,.1){\line(0,1){3.9}}
\put(2.5,.1){\line(0,1){3.9}}
\put(.1,1.5){\line(1,0){3.9}}
\put(.1,2.5){\line(1,0){3.9}} }

\savebox{\facs}(0,2){
\put(2,2) {\circle{4}}
\put(1.5,.1){\line(0,1){3.9}}
\put(2.5,.1){\line(0,1){3.9}}
\put(0.25,1.1){\line(2,1){3.5}}
\put(.25,2.9){\line(2,-1){3.5}} }

\savebox{\fact}(0,2){
\put(2,2) {\circle{4}}
\put(0.25,1.1){\line(2,1){3.5}}
\put(.25,2.9){\line(2,-1){3.5}}
\put(0,2){\line(1,0){4}}
\put(2,0){\line(0,1){4}} }

\savebox{\facu}(0,0){
\put(2,2) {\circle{4}}
\put(.2,1.5){\line(1,0){3.8}}
\put(2,0){\line(0,1){1.5}}
\put(.2,2.2){\line(1,0){3.8}}
\put(.4,3){\line(1,0){3.2}}}

\savebox{\facv}(0,0){
\put(2,2) {\circle{4}}
\put(.2,1.5){\line(1,0){3.8}}
\put(2,0){\line(0,1){1.5}}
\put(.2,2.2){\line(1,0){3.8}}
\put(2,2.2){\line(0,1){1.8}}}

\savebox{\facw}(0,0){
\put(2,2) {\circle{4}}
\put(0,2){\line(1,0){4}}
\put(2,0){\line(0,1){2.0}}
\put(1,.3){\line(0,1){1.7}}
\put(3,.25){\line(0,1){1.7}} }

\savebox{\facx}(0,0){
\put(2,2) {\circle{4}}
\put(.3,1.2){\line(1,0){3.6}}
\put(1.2,1.2){\line(0,1){1.6}}
\put(.3,2.8){\line(1,0){3.6}}
\put(2.8,1.2){\line(0,1){1.6}}}

\savebox{\facy}(0,0){
\put(2,2) {\circle{4}}
\put(0 ,2){\line(1,0){4}}
\put(1,.25){\line(0,1){1.8}}
\put(3,.25){\line(0,1){1.8}}
\put(.3,3){\line(1,0){3.4}} }

\begin{document}

\begin{titlepage}
\begin{flushright} { ~}  US-FT-18/98\\
hep-th/9812105\\ 
\end{flushright}
\vspace*{20pt}
\bigskip
\begin{center}
{\Large Vassiliev Invariants in the Context}
\vskip3mm
{\Large of Chern-Simons Gauge 
Theory\footnote{Invited lecture delivered by J. M. F. Labastida at the
workshop on ``New Developments in Algebraic Topology" held at Faro on July
13-15, 1998}}
\vskip 0.9truecm

\vspace{3pc}

{J. M. F. Labastida and  Esther P\'erez}

\vspace{1pc}

{\em  Departamento de F\'\i sica de Part\'\i culas,\\ Universidade de
Santiago de Compostela,\\ E-15706 Santiago de Compostela, Spain.\\}

\vspace{10pc}

{\large \bf Abstract}
\end{center} 

We summarize the progress made during the last few years on the study of
Vassiliev invariants from the point of view of perturbative Chern-Simons
gauge theory. We argue that this approach is the most promising one to
obtain a combinatorial universal formula for Vassiliev invariants. The
combinatorial expressions for the two primitive Vassiliev invariants of
order four, recently obtained in this context, are reviewed and rewritten
in terms of Gauss diagrams.

\end{titlepage}



Chern-Simons gauge theory has provided a very fruitful
context to study knot and link invariants. The multiple approaches inherent
to quantum field theory have been exploited to obtain different pictures for
the resulting invariants. Non-perturbative methods 
\cite{csgt,nbos,torus,king,martin,kaul} have established the connection of
Chern-Simons gauge theory with polynomial invariants as the Jones
polynomial \cite{jones} and its generalizations
\cite{homfly,kauffman,aku}. Perturbative methods
\cite{gmm,natan,vande,alla,torusknots,alts,thelinks,lcone} have provided
representations of Vassiliev invariants \cite{vass}. The purpose of
this lecture is to summarize the results obtained in recent years using the
latter methods.

Though it became clear some years ago that the terms of the perturbative
series expansion of Chern-Simons gauge theory were invariants of finite type
\cite{kont,barnatan,alla,bilin}, we had to wait until last
year to possess a field theory proof of this fact \cite{singular}. It was
shown in
\cite{singular},  that, after constructing 
gauge invariant operators for singular knots, the terms of the
perturbative series expansion of Chern-Simons gauge theory are invariants
of finite type. The proof is gauge independent and therefore the property
holds for any gauge-fixing. This result plus the fact that from a
non-perturbative point of view Chern-Simons gauge theory leads to the Jones
polynomial and its generalization constitutes a field theory proof of
Birman and Lin theorem \cite{bilin}.

Theories possessing gauge invariance, as Chern-Simons gauge theory, can be
studied  performing different gauge fixings. Vacuum expectation values of
gauge-invariant operators should be independent of the gauge fixing and they
can therefore be computed in different gauges.  Covariant gauges are 
simple to treat and its analysis in the case of perturbative Chern-Simons
gauge  theory has shown to lead to covariant  formulae for Vassiliev 
invariants
\cite{gmm,natan,alla,torusknots,alts}. These formulae involve 
multidimensional space and path integrals which, in general, are rather
involved to obtain the numerical value of Vassiliev invariants.
Non-covariant gauges seem  to lead to simpler formulae. However, the
subtleties inherent in non-covariant gauges \cite{leibrew} plague their
analysis with difficulties. The two non-covariant gauges more intensively
studied are the light-cone gauge and the temporal gauge
\cite{cata,cmartin,leibbrandt}. Both belong to  the general category of
axial gauges. In the light-cone gauge the resulting expressions for the
Vassiliev invariants turn out to be the ones involving Kontsevich integrals 
\cite{kont}. This was proven in \cite{lcone} and recently discussed in
\cite{kaucone}. The resulting expressions, although  simpler than the ones
appearing in covariant gauges,  are still too complicated to compute them
explicitly. In the temporal gauge one obtains much simpler expressions.
Actually, they do not involve integrations and are basically combinatorial
\cite{temporal}. Their explicit form up to order four has been presented in 
\cite{temporal}. 

Combinatorial expressions for Vassiliev invariants have
been seek since these invariants were formulated. To our knowledge, no
other method have been able to lead to this type of expressions up to order
four. An interesting combinatorial approach based on the use of Gauss
diagrams was introduced in \cite{arrgpo,willerton}. One of the goals of
this lecture is to show that the combinatorial expressions
obtained in \cite{temporal} can also be written in terms of Gauss diagrams.
However, our main goal is to argue that Chern-Simons gauge
theory is the most promising tool to build a  combinatorial universal
formula for Vassiliev invariants.

Non-covariant gauges  are difficult to treat in any quantum field theory
context \cite{leibrew}. Chern-Simons gauge theory is no exception to this.
However, in this case, due to the exact knowledge on the theory at our
disposal, it is known how the results obtained in a non-covariant gauge have
to be modified to find agreement with their covariant counterpart. In
computing vacuum expectation values of Wilson loops this turn out to be a
simple multiplicative factor \cite{lcone}, as first pointed out by
Kontsevich \cite{kont}. We will call this factor {\it Kontsevich factor}. A
similar phenomena seems to be present in the temporal gauge. In this case it
has been shown that the Kontsevich-like factor is not trivial and an
explicit expression for it has been conjectured \cite{temporal}. This
conjecture has been proved up to order four. Understanding the origin of
the Kontsevich factor one could gain some insight on some of the
general problems inherent to non-covariant gauges.

We will begin reviewing the salient facts of the analysis of the
perturbative series expansion of the vacuum expectation value of a Wilson
loop in the temporal gauge carried out in \cite{temporal}. Given a knot $K$
and one of its regular knot projections, ${\cal K}$, on the $x_1,x_2$-plane
which is a Morse knot in the $x_1$ and $x_2$ directions, one possesses a
perturbative series expansion for the vacuum expectation value of the
corresponding Wilson loop:
\beq \langle W({K},G)\rangle = \langle W({\k},G)\rangle_{{\rm temp}}
\times  \langle W(U,G)\rangle^{b(\k)}, 
\label{global} 
\eeq
being,
\beq
{1 \over d} \langle W({K},G)\rangle = 1 + \sum_{i=1}^{\infty} v_i(K) x^i,
\label{expansiona}
\eeq
and,
\beq
{1 \over d} \langle W({\k},G)\rangle_{{\rm temp}} = 1 + \sum_{i=1}^{\infty} 
\hat v_i(\k)  x^i.
\label{expansionb}
\eeq
In these expressions $x$ denotes the inverse of the Chern-Simons coupling
constant, $x=2\pi i/k$, $G$ the gauge group, and $d$ the dimension of the
representation carried by the Wilson loop. The function $b({\cal K})$ is
the exponent of the Kontsevich factor, which has been conjectured to be
\cite{temporal},
\beq
b(\k) = {1\over 12} (n_{x_1}+n_{x_2}),
\label{hipotesis}
\eeq
where $n_{x_1}$ and $n_{x_2}$ are the critical points of the regular
projection ${\cal K}$ in both, the $x_1$ and the $x_2$ directions. In
(\ref{global}) $U$ denotes the unknot and  $\langle W({\k},G)\rangle_{{\rm
temp}}$ is the vacuum expectation of the Wilson line corresponding to the
regular projection ${\cal K}$ as computed perturbatively in the temporal
gauge with the standard Feynman rules of the theory. Notice that though each
of the factors on the right hand side of (\ref{global}) depends on the
regular projection chosen, the left hand side does not. While the
coefficients $v_i(K)$ of the series (\ref{expansiona}) are Vassiliev
invariants the coefficients $\hat v_i(K)$  of (\ref{expansionb}) are not. 
The latter depend on the regular projection chosen.

\begin{figure}
\centerline{\epsffile{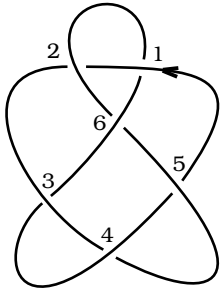}}
\caption{Example of a knot projection.}
\label{labels}
\end{figure}

An explicit combinatorial  form (no integrals left) of the coefficients
$\hat v_i(K)$  in (\ref{expansionb}) would lead to a universal combinatorial
formula for Vassiliev invariants. Unfortunately, this has not been obtained
yet at all orders. Only part of the contributions entering $\hat v_i(K)$
have been explicitly written at all orders. These are the {\it kernels}
introduced in \cite{temporal}. The kernels are quantities which depend on
the knot projection chosen and therefore are not knot invariants. However,
at a given order $i$ a kernel differs from an invariant of type $i$ by
terms that vanish in signed sums of order $i$. The kernel
contains the part of a Vassiliev which is the last in becoming zero when
performing signed sums, in other words, a kernel vanishes in signed sums of
order $i+1$ but does not in signed sums of order $i$. In some sense the
kernel represents the most fundamental part of a Vassiliev invariant, \ie,
the part that survives a maximum number of signed sums. Kernels plus the
structure of the perturbative series expansion seem to contain enough
information to reconstruct the full Vassiliev invariants. This was shown in
\cite{temporal} up to order four. The results obtained there will be
presented below and rewritten in a more compact form.

The expression for the kernels results after considering only the
simplest part of the propagator of the gauge field in the temporal gauge.
This part involves a double delta function and therefore all the integrals
can be performed. The result is a combinatorial expression in terms of
crossing signatures after distributing propagators among all the
crossings. The general expression can be written in a universal form much
in the spirit of the universal form of the Kontsevich integral \cite{kont}.
Let us consider a knot $K$ with a regular knot projection ${\cal K}$
containing $n$ crossings. Let us choose a base point on ${\cal K}$ and let
us label the $n$ crossings by $1,2,\dots,n$ as we pass for
first time through each of them when traveling along ${\cal K}$ starting
at the base point. The universal expression for the kernel associated to
${\cal K}$ has the form:
\beq 
{\cal N}(\k) = \sum_{k=0}^\infty\Bigg(
\sum_{m=1}^k \sum_{p_1,\dots,p_m =1\atop p_1+\cdots+p_m=k}^k
\sum_{i_1,\dots,i_m=1\atop i_1 < \cdots < i_m}^n
{\epsilon_{i_1}^{p_1} \cdots \epsilon_{i_m}^{p_m} \over
(p_1!\cdots p_m!)^2}\sum_{\sigma_{1},\dots,\sigma_{m} \atop
\sigma_{1}\in P_1,\dots,\sigma_{m}\in P_m}
{\cal T}(i_1,\sigma_{1};\dots;i_m,\sigma_{m})\Bigg).
\label{nucleos}
\eeq
In this equation $P_m$ denotes the permutation group of $p_m$ elements. The
factors in the innest sum, ${\cal T}(i_1,\sigma_{1};\dots;i_m,\sigma_{m})$,
are group factors which are computed in the following way:
given a set of crossings, $i_1, \dots, i_{m}$, and a set of permutations,
$\sigma_1,\dots,\sigma_m$, with $\sigma_1\in P_1,\dots,\sigma_m\in P_m$,
the corresponding group factor
${\cal T}(i_1,\sigma_{1};\dots;i_m,\sigma_{m})$ is the result of taking a
trace over the  product
of group generators which is obtained after assigning $p_1,\dots,p_m$ group
generators to the crossings $i_1, \dots, i_{m}$ respectively, and placing
each set of group generators in the order which results after traveling
along the knot starting from the base point. The first time that one
encounters a crossing
$i_j$ a product of
$p_j$ group generators is introduced; the  second
time the product is similar, but with the indices rearranged according to 
the permutation $\sigma_j\in P_j$.

\begin{figure}
\centerline{\epsffile{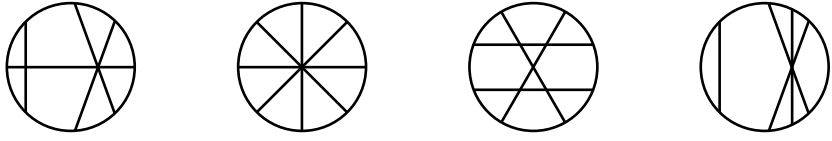}}
\caption{Chord diagrams corresponding to group factors.}
\label{exgroup}
\end{figure}

In order to clarify the content of (\ref{nucleos}) we will  work out an 
example. Let us consider the knot projection shown in fig. \ref{labels} and
let us concentrate  on some of the fourth order contributions, $k=4$. The
knot projection under consideration has $n=5$ crossings. We will consider,
for example, terms with $m=3$, and, $p_1=2$, $p_2=1$ and $p_3=1$. Since in
this case the permutation groups $P_2$ and $P_3$  contain
only the identity element, $1$, the form of the kernel is:
\beq {1\over
(2!)^2} \sum_{ i_1,i_2,i_3 =1 \atop i_1 < i_2 < i_3}^6 \epsilon_{i_1}^2
\epsilon_{i_2} \epsilon_{i_3} {\cal T}(i_1,\sigma_1;i_2,1;i_3,1),
\label{ejem}
\eeq
where  $\sigma_1\in P_1$, being $P_1$ the permutation group of 2 elements.
Examples of the group factors entering this expression are:
\bear
{\cal T}(1,\sigma_1;2,1;3,1) &=& \tr ( T^{b_1}T^{b_2}T^{a_1}T^{a_2}T^{a_1}
T^{\sigma_1(b_1)}T^{\sigma_1(b_2)}T^{a_2}), \nonumber \\
{\cal T}(1,\sigma_1;3,1;5,1) &=& \tr ( T^{b_1}T^{b_2}T^{a_1}T^{a_2}
T^{\sigma_1(b_1)}T^{\sigma_1(b_2)}T^{a_1}T^{a_2}), \nonumber \\
{\cal T}(2,\sigma_1;3,1;6,1) &=& \tr ( T^{b_1}T^{b_2}T^{a_1}T^{a_2}
T^{\sigma_1(b_1)}T^{\sigma_1(b_2)}T^{a_2}T^{a_1}), \nonumber \\
{\cal T}(3,\sigma_1;4,1;6,1) &=& \tr ( T^{b_1}T^{b_2}T^{a_1}T^{a_2}
T^{a_2}T^{\sigma_1(b_1)}T^{\sigma_1(b_2)}T^{a_1}), \nonumber \\
\label{masejem}
\eear
where we have used the labels specified in fig. \ref{labels}. Group factors 
can be represented by chord diagrams. For example if one chooses
$\sigma_1=(12)$ the four chord diagrams corresponding to the group factors
in (\ref{masejem})  are the ones pictured in fig. \ref{exgroup}. The
kernels are independent of the base point chosen for ${\cal K}$. 

The universal formula (\ref{nucleos}) for the kernels can be written in a
more useful way collecting all the coefficients multiplying a given group
factor. The group factors can be labeled by chord diagrams. At  order
$k$ one has a term for each of the inequivalent chord diagrams with $k$
chords. Denoting chord diagrams by $D$, equation (\ref{nucleos}) can be
written as:
\beq 
{\cal N}(\k) = \sum_{D} N_D(\k) D,
\label{masnucleos}
\eeq
where the sum extends to all inequivalent chord diagrams.
Our next task is to derive from (\ref{nucleos})  the general form of the
kernels $N_D(\k)$. The concept of kernel can be
extended to include singular knots by considering signed sums of
(\ref{masnucleos}), or, following
\cite{singular}, introducing vacuum expectation values of the operators for
singular knots. If $\k^j$ denotes a regular projection of a knot $K^j$ with
$j$ simple singular crossings or double points, the corresponding universal
form for the kernel possesses an expansion similar to  (\ref{masnucleos}):
\beq 
{\cal N}(\k^j) = \sum_{D} N_D(\k^j) D.
\label{sinnucleos}
\eeq
The general results about singular knots proved in \cite{singular} lead to
two important features for (\ref{sinnucleos}). On the one hand,
finite type implies that $N_D(\k^j)=0$ for chord diagrams $D$ with more
than $j$ chords. On the other hand, $N_D(\k^j)=2^j \delta_{D,D({\cal
K}^j)}$, where $D({\cal K}^j)$ is the configuration corresponding to
the singular knot projection ${\cal K}^j$. As observed above, kernels
constitute the part of a Vassiliev invariant which survives a maximum
number of signed sums.

\begin{figure}
\centerline{\epsffile{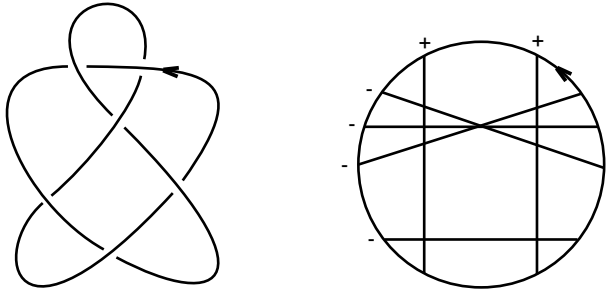}}
\caption{A regular knot projection and its corresponding Gauss diagram.}
\label{seis}
\end{figure}

    To compute $N_D(\k)$ we will introduce first the notion of the set of
labeled chord subdiagrams of a given chord diagram. We will denote this set
by $S_D$. This set is made out of a selected set of labeled chord diagrams
that we now define.

\vskip0.3cm
A {\it labeled chord diagram} of order $p$ is a chord diagram with $p$
chords and a set of positive integers $k_1,k_2,\dots,k_p$, which will be
called labels, such that each chord has one of these integers attached.
\vskip0.3cm

The set $S_D$ is made out of labeled chord diagrams which satisfy two
conditions. These conditions are fixed by the form of the series entering
the kernels (\ref{nucleos}). We will call the elements of $S_D$ labeled
chord subdiagrams of the chord diagram $D$. They are defined as follows.

\vskip0.3cm
A {\it labeled chord subdiagram} of a chord diagram $D$ with $k$ chords is a
labeled chord diagram of order $p$ with labels $k_1,k_2,\dots,k_p$, $p\leq
k$, such that the following two conditions are satisfied: 

{\sl a)}
$k_1+k_2+\cdots+k_p=k$; 

{\sl b)} there exist  elements $\sigma_1 \in
P_{k_1},\sigma_2\in P_{k_2},\dots,\sigma_p\in P_{k_p}$ of the permutation
groups
$P_{k_1},P_{k_2},\dots,P_{k_p}$ such that, after replacing the $j$-th chord
diagram by $k_j$ chords arranged according to the permutation
$\sigma_j$, for $j=1,\dots,p$, the resulting chord diagram is homeomorphic
to $D$. The number of ways that permutations  
$\sigma_1 \in P_{k_1},\sigma_2\in P_{k_2},\dots,\sigma_p\in P_{k_p}$ can be
chosen is called the multiplicity of the labeled chord subdiagram. We will
denote the multiplicity of a given labeled chord subdiagram, $s\in S_D$, by
$m_D(s)$.
\vskip0.3cm

The chord diagram $D$ itself can be
regarded as a labeled chord subdiagram such that its labels, or positive
integers attached to its chords, are 1. It has multiplicity 1. All
the elements of $S_D$ except
$D$ have a number of chords smaller than the number of chords of $D$. Not
all labeled chord diagrams are subdiagrams of $D$. However, given a labeled
chord diagram with labels  $k_1,k_2,\dots,k_p$ there can be different sets
of permutations leading to $D$. The number of these different sets is the
multiplicity introduced above. The elements of the sets
$S_D$ for all chord diagrams $D$ up to order four
which do not have disconnected subdiagrams are the following:
\begin{equation}
\vbox{\begin{eqnarray*}
\cdosii &\longrightarrow & \hskip0.3cm \cdosii \hbox{\hskip-0.4cm}, 
\cunodos \\
\vbox{\vskip0.9cm}
\ctresiii &\longrightarrow & \hskip0.3cm \ctresiii \hbox{\hskip-0.4cm}, 
\cdosiidosuno
\hbox{\hskip-0.4cm}, 2
\cunotres
\\
\vbox{\vskip0.9cm}
\ctresiv &\longrightarrow & \hskip0.3cm \ctresiv \hbox{\hskip-0.4cm},
 \cdosiidosuno
\hbox{\hskip-0.4cm},  \cunotres
\\
\vbox{\vskip0.9cm}
\ccuavi &\longrightarrow & \hskip0.3cm \ccuavi \hbox{\hskip-0.4cm}, 
\ctresiiidos \hbox{\hskip-0.4cm}, \cdosiitresuno
\hbox{\hskip-0.4cm},  2 \cunocuatro
\\
\vbox{\vskip0.9cm}
\ccuavii &\longrightarrow & \hskip0.3cm \ccuavii \hbox{\hskip-0.4cm}, 
 2 \cunocuatro
\\
\vbox{\vskip0.9cm}
\ccuaviii &\longrightarrow & \hskip0.3cm \ccuaviii \hbox{\hskip-0.4cm}, 
\ctresiiidos \hbox{\hskip-0.4cm}, 2\cdosiitresuno
\hbox{\hskip-0.4cm},  4 \cunocuatro
\\
\vbox{\vskip0.9cm}
\ccuaix &\longrightarrow & \hskip0.3cm \ccuaix \hbox{\hskip-0.4cm}, 
\ctresiiidosbis \hbox{\hskip-0.4cm}, 2\cdosiidosdos
\hbox{\hskip-0.4cm},   \cunocuatro
\\
\vbox{\vskip0.9cm}
\ccuax &\longrightarrow & \hskip0.3cm \ccuax \hbox{\hskip-0.4cm}, 
\ctresivdos \hbox{\hskip-0.4cm}, \ctresiiidosbis
\hbox{\hskip-0.4cm}, 2\cdosiidosdos \hbox{\hskip-0.4cm},
2\cdosiitresuno \hbox{\hskip-0.4cm}, 3\cunocuatro
\\
\vbox{\vskip0.9cm}
\ccuaxi &\longrightarrow & \hskip0.3cm \ccuaxi \hbox{\hskip-0.4cm}, 
\ctresivdos \hbox{\hskip-0.4cm}, \cdosiidosdos \hbox{\hskip-0.4cm},
\cdosiitresuno \hbox{\hskip-0.4cm}, \cunocuatro
\end{eqnarray*}}
\end{equation}
The numbers accompanying each labeled chord subdiagram denote their
multiplicity. When no number is attached to a chord of a labeled chord
diagram it should be understood that the corresponding label is 1.

\begin{figure}
\centerline{ \epsffile{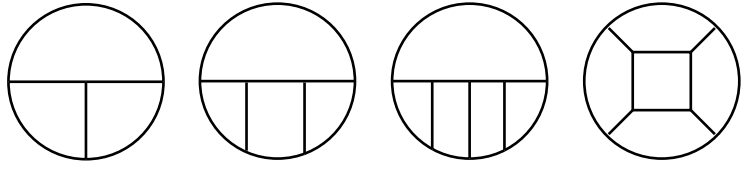}}
\caption{Basis of primitive Vassiliev invariant up to order four.}
\label{base}
\end{figure}

In order to write our final expression for the kernels we need to recall the
notion of Gauss diagram. Given a regular projection ${\cal K}$ of a knot
$K$ we can associate to it its Gauss diagram $G({\cal K})$. The regular
projection ${\cal K}$ can be regarded as a generic immersion of a circle
into the plane enhanced by information on the crossings. The Gauss diagram
$G({\cal K})$ consists of a circle together with the preimages of each
crossing of the immersion connected by a chord. Each chord is equipped with
the sign of the signature of the corresponding crossing. An example of
Gauss diagram has been pictured in fig. \ref{seis}. Gauss diagrams are
useful because they allow to keep track of the sums involving the crossings
which enter in (\ref{nucleos}) in a very simple form. Let us consider a
chord diagram $D$ and one of its labeled chord subdiagrams $s\in S_D$. Let
us assume that $s$ has $p$ chords and labels
$k_1,k_2,\cdots,k_p$. We define the product,
\beq
\langle s, G({\cal K}) \rangle,
\label{producto}
\eeq 
as the sum over all the embeddings of $s$ into $G({\cal K})$, each
one weighted by a factor,
\beq
{\epsilon_1^{k_1} \epsilon_2^{k_2}\cdots \epsilon_p^{k_p} \over
(k_1! k_2! \cdots k_p!)^2},
\label{pesos}
\eeq
where $\epsilon_1, \epsilon_2, \dots, \epsilon_p$ are the signatures
of the chords of $G({\cal K})$ involved in the embedding. Using
(\ref{producto})  the kernels $N_D(\k)$ entering (\ref{masnucleos}) can be
written as,
\beq
 N_D(\k) = \sum_{s\in S_D} m_D(s) \langle s, G({\cal K}) \rangle,
\label{laformula}
\eeq
where $m_D(s)$ denotes the multiplicity of the labeled subdiagram $s\in
S_D$ relative to the chord diagram $D$.

The product (\ref{producto}) possesses important properties. First, it is
independent of the base point chosen for the regular projection ${\cal K}$
and, correspondingly, for the Gauss diagram $G(\k)$. Second, it is of finite
type. This means that if $s$ has $j$ chords, the result of computing a
signed sum of order higher than $j$ is zero. Recall that signed sums of
order $k$ are used to define quantities associated to singular knot
projections with $k$ double points, as the ones entering
(\ref{masnucleos}). A signed sum of order $k$ contains $2^k$ terms which
correspond to the possible ways of resolving $k$ double points into 
overcrossings and undercrossings. Each one has a sign which corresponds to
the product of the signatures of the crossings involved in the $k$ double
points. If $s$ is a labeled chord diagram with $j$ chords and all its labels
take value one, the order-$j$ signed sum is $2^j$ if the configuration of
the singular projection with $j$ double points associated to such a sum
corresponds to the chord diagram $s$; otherwise its value is zero. This
fact leads to the result mentioned above stating that:
\beq
N_D(\k^j)=2^j \delta_{D,D({\cal
K}^j)},
\label{manzana}
\eeq
where $D({\cal K}^j)$ is the configuration corresponding to the
singular knot projection associated to the signed sum. Of course, the
product (\ref{producto}) vanishes if the number of chords of $s$ is bigger
than the number of chords of the Gauss diagram $G(\k)$.

The products (\ref{producto}) can be regarded as quantities of finite type
associated to Gauss diagrams $G$ whether or not these correspond to a
regular projection of a knot. Gauss diagrams can be studied as abstract
objects characterized by chord diagrams with signs assigned to their
chords. It is clear that in such a general context the quantities
$ \langle s , G \rangle $, as defined in (\ref{producto}), are of finite
type. In other words, if $s$ has $j$ chords and $G$ is an abstract Gauss
diagram, the product $\langle s , G \rangle $ vanishes under signed sums of
order higher than $j$. This observation leads to conjecture that the
product (\ref{producto}) might play an interesting role in the theory of
virtual knots \cite{virkau,virgpo}.

The terms $\langle s, G({\cal K}) \rangle$ entering (\ref{laformula}) are
related to the quantities $\chi({\cal K})$ defined in \cite{temporal}. It
is straightforward to obtain the following relations:
\beq
\vbox{\hskip-5cm
\vbox{\begin{eqnarray*}
\langle  \cunoj \hbox{\hskip-0.4cm}, G({\cal K}) \rangle &=& {1\over
(j!)^2}\chi_1({\cal K}),
\hbox{\hskip0.3cm} j \hbox{\hskip0.2cm} \hbox{\rm odd,} \nonumber \\
\vbox{\vskip0.9cm}
\langle \cdosii \hbox{\hskip-0.4cm}, G({\cal K}) \rangle
&=& 
\chi_2^A({\cal K}), \nonumber \\ 
\vbox{\vskip0.9cm}
\langle \cdosiidosdos \hbox{\hskip-0.4cm}, G({\cal K})
\rangle &=&  {1\over 16} \chi_2^C({\cal K}), \nonumber \\ 
\vbox{\vskip0.9cm}
\langle \ctresiii \hbox{\hskip-0.4cm}, G({\cal K})
\rangle &=& 
 \chi_3^B({\cal K}), \nonumber \\ 
\vbox{\vskip0.9cm}
\langle \ctresiiidos \hbox{\hskip-0.4cm}, G({\cal K})
\rangle &=&  {1\over 4} \chi_3^D({\cal K}), \nonumber \\ 
\vbox{\vskip0.9cm}
\langle \ccuaxi \hbox{\hskip-0.4cm}, G({\cal K}) \rangle
&=& 
\chi_4^A({\cal K}), \nonumber \\ 
\vbox{\vskip0.9cm}
\langle \ccuaix \hbox{\hskip-0.4cm}, G({\cal K}) \rangle
&=& 
\chi_4^C({\cal K}), \nonumber \\ 
\vbox{\vskip0.9cm}
\langle \ccuaviii \hbox{\hskip-0.4cm}, G({\cal K})
\rangle &=& 
\chi_4^E({\cal K}), \nonumber 
\vbox{\vskip0.9cm}
\end{eqnarray*}}

\vskip-10.3cm

\hskip3.2cm
\vbox{\begin{eqnarray*}
\langle \cunoj \hbox{\hskip-0.4cm}, G({\cal K}) \rangle &=& {1\over
(j!)^2}n({\cal K}),
\hbox{\hskip0.3cm} j \hbox{\hskip0.2cm} \hbox{\rm even,} \nonumber \\
\vbox{\vskip0.9cm}
\langle \cdosiidosuno \hbox{\hskip-0.4cm}, G({\cal K}) \rangle &=& 
{1\over 4} \chi_2^B({\cal K}), \nonumber \\ 
\vbox{\vskip0.9cm}
\langle \ctresiv \hbox{\hskip-0.4cm}, G({\cal K}) \rangle
&=& 
 \chi_3^A({\cal K}), \nonumber \\ 
\vbox{\vskip0.9cm}
\langle \ctresivdos \hbox{\hskip-0.4cm}, G({\cal K})
\rangle &=&  {1\over 4} \chi_3^C({\cal K}), \nonumber \\ 
\vbox{\vskip0.9cm}
\langle \ctresiiidosbis \hbox{\hskip-0.4cm}, G({\cal K})
\rangle &=&  {1\over 4} \chi_3^E({\cal K}), \nonumber \\ 
\vbox{\vskip0.9cm}
\langle \ccuax \hbox{\hskip-0.4cm}, G({\cal K}) \rangle
&=& 
\chi_4^B({\cal K}), \nonumber \\ 
\vbox{\vskip0.9cm}
\langle \ccuavi \hbox{\hskip-0.4cm}, G({\cal K}) \rangle
&=& 
\chi_4^D({\cal K}), \nonumber \\ 
\vbox{\vskip0.9cm}
\langle \ccuavii \hbox{\hskip-0.4cm}, G({\cal K}) \rangle
&=& 
\chi_4^F({\cal K}).  \nonumber 
\end{eqnarray*}}
}\label{lalista}
\eeq
Notice that in the second relation $n({\cal K})$ denotes the number of
crossings of the regular projection ${\cal K}$. The rest of the quantities
on the right hand side of (\ref{lalista}) were defined in \cite{temporal}.

\begin{figure}
\centerline{ \epsffile{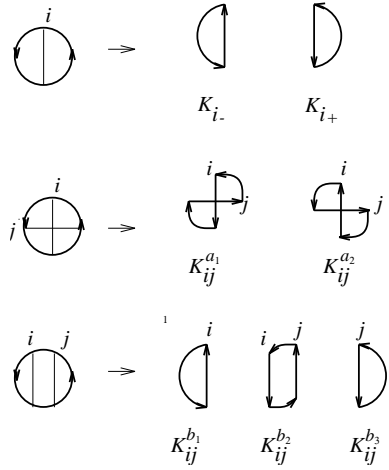}}
\caption{Splitting a knot into other knots.}
\label{subknots}
\end{figure}

In \cite{temporal} we were
able to express all the Vassiliev invariants up to order four in terms of
these quantities and the crossing signatures. The strategy was to start with
the kernels (\ref{laformula}) and exploit the properties of the perturbative
series expansion of Chern-Simons gauge theory. A special role in the
construction was played by the  factorization theorem proved in
\cite{factor}. At orders two and three there is only one primitive Vassiliev
invariant. We will make the same choice of basis as in \cite{temporal}. The
diagrams associated to them are the first two in fig. \ref{base}. The two
primitive Vassiliev invariants turn out to be, at second order,
\beq
\alpha_{21}(K) = \alpha_{21}(U) +  \langle \cdosii \hbox{\hskip-0.4cm}, \bar
G(\k) \rangle,
\label{primtwoc}
\eeq
while, at third order,
\beq
\alpha_{31} (K) = \langle \ctresiii \hbox{\hskip-0.4cm} + \ctresiv
\hbox{\hskip-0.4cm} + 2 \cdosiidosuno
\hbox{\hskip-0.4cm},  G(\k) \rangle
 - \sum_{i=1}^n \, \epsilon_i(\k) \Big[ \langle  \cdosii
\hbox{\hskip-0.4cm}, G(\alpha(\k)) \rangle \Big]_i.
\label{primthreeb}
\eeq
Several comments are in order to explain the quantities entering  these
expressions. In (\ref{primtwoc}) $\alpha_{21}(U)$ stands for the value of
the invariant $\alpha_{21}$ for the unknot. In the first equation
the bar denotes that the product has to be taken on $G(\k)$ and then
substract its value for the ascending diagram.
In general a bar over a quantity
$L(\k)$ indicates that the same quantity for the ascending diagram has to
be subtracted, \ie:
\beq
\bar L(\k) = L(\k) - L(\alpha(\k))
\label{limon}
\eeq
where
$\alpha({\cal K})$ denotes the standard ascending diagram of ${\cal K}$.
The ascending diagram of a knot projection is defined as the diagram
obtained by switching, when traveling along the knot from a base point, 
all the undercrossings to overcrossings. In (\ref{primthreeb}) the sum is
over all crossings $i$, $i=1,\dots,n$, and $\epsilon_i(\k)$ denotes the
corresponding signature. The square brackets $[$ $]_i$ enclosing a quantity
$L(\k)$ denote:
\beq
 \Big[  L(\k) \Big]_i = L(\k) - L(\k_{i_+}) -L(\k_{i_-}),
\label{platano}
\eeq
where the regular projection diagrams $\k _{i_+}$ and
$\k _{i_-}$ are the ones which result after the splitting of $\k$ at the
crossing point $i$ as shown in the first row of fig. \ref{subknots}. It is
clear from the list (\ref{lalista}) that these two invariants can be
written in terms of the products (\ref{producto}) and the crossing
signatures.

Combinatorial expressions for the two primitive invariants at order four
have been presented in \cite{temporal}. Their construction is based on the
use of the kernels (\ref{laformula}) and the properties of the perturbative
series expansion. As in the case of previous orders, these invariants are
expressed in terms of the products (\ref{producto}) and the crossing
signatures. Their form is more complicated than the ones at lower orders.
They turn out to be:
\bear
&& {\hskip -0.5cm} \alpha_{42}(K) = \alpha_{42}(U) +
\langle 7 \ccuaxi \hbox{\hskip-0.4cm} + 5 \ccuax \hbox{\hskip-0.4cm} + 4
\ccuaix \hbox{\hskip-0.4cm} + 2 \ccuaviii \hbox{\hskip-0.4cm} + \ccuavi
\hbox{\hskip-0.4cm} + \ccuavii \hbox{\hskip-0.4cm}  \nonumber \\ && 
\nonumber \\ && 
\hbox{\hskip6.18cm} + 8
\ctresivdos
\hbox{\hskip-0.4cm} + 2 \ctresiiidos \hbox{\hskip-0.4cm} + 8 \ctresiiidosbis
\hbox{\hskip-0.4cm} +{1\over 6} \cdosii \hbox{\hskip-0.4cm},
\bar G(\k) \rangle  \nonumber \\ && +
\sum_{i,j\in {\cal C}_a\atop i>j} \bar\epsilon_{ij}(\k)\Bigg(
\Big[\langle \cdosii \hbox{\hskip-0.4cm}, G(\alpha(\k))\rangle \Big]_{ij}^a
-2
\Big[\langle \cdosii \hbox{\hskip-0.4cm}, G(\alpha(\k))\rangle\Big]_i - 2
\Big[\langle \cdosii \hbox{\hskip-0.4cm}, G(\alpha(\k))\rangle\Big]_j
\Bigg)
\nonumber \\ && +
\sum_{i,j\in {\cal C}_b\atop i>j} \bar\epsilon_{ij}(\k)\Bigg(
\Big[\langle \cdosii \hbox{\hskip-0.4cm}, G(\alpha(\k))\rangle\Big]_{ij}^b -
\Big[\langle \cdosii \hbox{\hskip-0.4cm}, G(\alpha(\k))\rangle\Big]_i - 
\Big[\langle \cdosii \hbox{\hskip-0.4cm}, G(\alpha(\k))\rangle\Big]_j
\Bigg),  \nonumber \\ 
\label{primfourm}
\eear
and,
\bear
&& {\hskip -0.7cm} \alpha_{43}(K) = \alpha_{43}(U) + 
\langle  \ccuaxi \hbox{\hskip-0.4cm} + \ccuax \hbox{\hskip-0.4cm} + 
\ccuaix \hbox{\hskip-0.4cm} + 2 \ctresiiidosbis \hbox{\hskip-0.4cm}
- {1\over 6} \cdosii \hbox{\hskip-0.4cm}, \bar G(\k) \rangle
\nonumber \\ && +
\sum_{i,j\in {\cal C}_a\atop i>j} \bar\epsilon_{ij}(\k)\Bigg(
\Big[\langle \cdosii \hbox{\hskip-0.4cm},
G(\alpha(\k))\rangle\Big]_{ij}^a-\Big[\langle \cdosii \hbox{\hskip-0.4cm},
G(\alpha(\k))\rangle\Big]_i - 
\Big[\langle \cdosii \hbox{\hskip-0.4cm}, G(\alpha(\k))\rangle\Big]_j
\Bigg). \nonumber \\
\label{primfourn}
\eear
In these expressions the explicit dependence on the signatures appears in
the quantities $\bar\epsilon_{ij}(\k)$ which are:
\beq
\bar\epsilon_{ij}(\k) = \epsilon_{ij}(\k) - \epsilon_{ij}(\alpha(\k))=
\epsilon_{i}(\k)\epsilon_{j}(\k) -
\epsilon_{i}(\alpha(\k))\epsilon_{j}(\alpha(\k)).
\label{lasepsilons}
\eeq
The sums in which these products are involved are over double splittings of
the knot projection $\k$ at the crossings $i$ and $j$. There are two ways of
carrying out these double splittings, depending on the configuration
associated to the crossings $i$ and $j$. These are shown in the second and
third rows of fig.
\ref{subknots}. In the first one the regular projection $\k$ is split 
into two while in the second
one it is split into three. Splittings of the first type build the set
${\cal C}_a$. The ones of the second type build ${\cal C}_b$. While only
the first one is involved in the invariant $\alpha_{43}$, both appear
in $\alpha_{42}$. The new quantities entering the sums are:
\bear
\Big[L(\k)\Big]_{ij}^a &=&
L(\k) - L(\k_{ij}^{a_1}) -  L(\k_{ij}^{a_2}),
\nonumber \\
\Big[L(\k)\Big]_{ij}^b &=&
L(\k) - L(\k_{ij}^{b_1}) -  L(\k_{ij}^{b_2}) -  L(\k_{ij}^{b_3}),
\label{sandia}
\eear
where $\k_{ij}^{a_1},\k_{ij}^{a_2},\k_{ij}^{b_1},\k_{ij}^{b_2}$ and
$\k_{ij}^{b_3}$ are the knot projections which originate after a double
splitting of $\k$, as denoted in fig.
\ref{subknots}. As in previous orders, in the  expressions
(\ref{primfourm}) and (\ref{primfourn}), the quantities
$\alpha_{42}(U)$ and
$\alpha_{43}(U)$ correspond to the value of these invariants for the
unknot. It has been proved in \cite{temporal} that the combinatorial
expressions for $\alpha_{42}(K)$ and
$\alpha_{43}(K)$ in (\ref{primfourm}) and (\ref{primfourn}) are invariant
under Reidemeister moves.

\begin{figure}
\centerline{ \epsffile{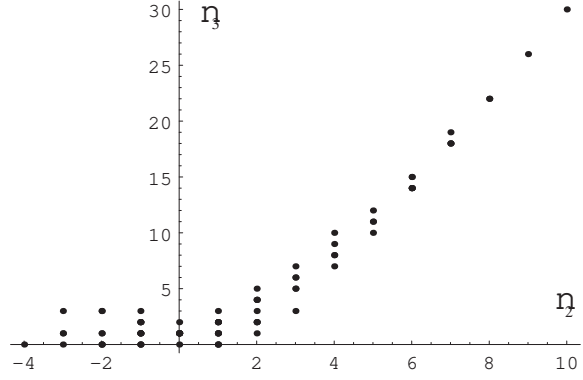}}
\caption{Plot of the absolute value of the third-order Vassiliev invariant
$\nu_3$ versus the one of order two, $\nu_2$, for all prime knots up to nine
crossings.}
\label{tres}
\end{figure}

\begin{figure}
\centerline{\epsffile{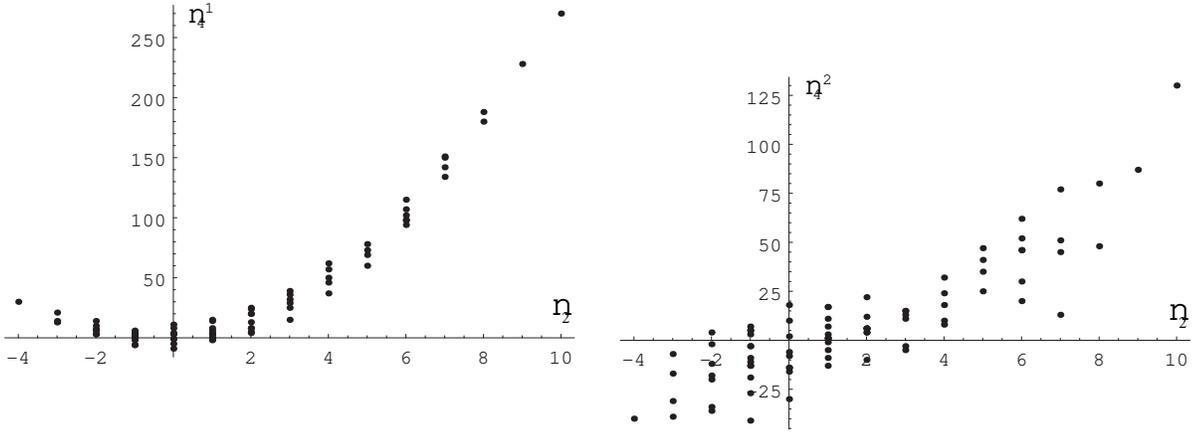}}
\caption{Plots of the two fourth-order Vassiliev invariants $\nu_4^1$ and
$\nu_4^2$ versus the second order one $\nu_2$, for all prime knots up to
nine crossings.}
\label{cuatro}
\end{figure}

Vassiliev invariants constitute vector spaces and their normalization can
be chosen in such a way that they are integer-valued. Once their value for
the unknot has been subtracted off they can be presented in many
basis in which they are integers. We will chose here a  particular basis in
which the numerical values for the invariants up to order four are rather
simple:
\bear
\nu_{2}(K) &=& {1\over 4} \tilde\alpha_{21}(K), \nonumber \\
\nu_{3}(K) &=& {1\over 8} \tilde\alpha_{31}(K), \nonumber \\
\nu_{4}^1(K) &=& {1\over 8} (\tilde\alpha_{42}(K) + \tilde\alpha_{43}(K)),
\nonumber \\
\nu_{4}^2(K) &=& {1\over 4} (\tilde\alpha_{42}(K) - 5 \tilde\alpha_{43}(K)),
\nonumber \\
\label{basica}
\eear
where the tilde indicates that the value for the unknot has been
subtracted, \ie, $\tilde\alpha_{ij}(K)=\alpha_{ij}(K)-\alpha_{ij}(U)$. In
Tables 1 and 2 we have collected the value of the Vassiliev invariants
(\ref{basica}) for all prime knots up to nine crossings. Notice that we
could have chosen a basis where all the values for the trefoil knot are 1 
just redefining $\nu_{4}^1(K)$ into $\nu_{4}^1(K)-2 \nu_{4}^2(K)$. We have
no done so because $\nu_{4}^1(K)$, as defined in (\ref{basica}), has a
simple shape when plotted versus $\nu_2(K)$. Actually, the resulting shape
has features similar to the shape which results after plotting $\nu_3(K)$
versus
$\nu_2(K)$. In fig.
\ref{cuatro}  we present  $\nu_{4}^1(K)$ and  $\nu_{4}^2(K)$ versus 
$\nu_2(K)$. These should be compared to the plot of the absolute value of 
$\nu_3(K)$ versus $\nu_2(K)$ depicted in fig. \ref{tres}. The similar
behavior observed for $|\nu_3(K)|$ and $\nu_{4}^2(K)$ is expected from
their general form for torus knots. As it was shown in \cite{torusknots} and
\cite{simon}, for a torus knot characterized by two coprime integers $p$ and
$q$ these invariants are the following polynomials in $p$ and $q$:
\bear
\nu_2(p,q) &=& {1\over 24}  (p^2-1)(q^2-1) \nonumber \\
\nu_3(p,q) &=&  {1\over 144} (p^2-1)(q^2-1)pq \nonumber \\
\nu_4^1(p,q) &=& {1\over 288} (p^2-1)(q^2-1)p^2q^2 \\
\nu_4^2(p,q) &=& {1\over 720} (p^2-1)(q^2-1)(2p^2q^2-3p^2-3q^2-3) \nonumber
\label{toros}
\eear
The explicit expression of Vassiliev invariants as polynomials in $p$ and
$q$ is known up to order six \cite{torusknots}. Of course, up to order four
their value agree with the ones computed explicitly from equations
(\ref{primfourm}) and (\ref{primfourn}), as it can be checked explicitly
from the tables collected below. The only torus knots up to nine crossings
are $3_1$,
$5_1$, $7_1$, $8_{19}$ and $9_1$, whose associated coprime integers are
(3,2), (5,2), (7,2), (4,3) and (9,2), respectively.

 It would be desirable to write the invariants in such a way that
signatures and split sums do not appear. Even better would be to possess
expressions where terms involving ascending diagrams are not present. It is
not known if this is possible even  for the few orders in which
combinatorial expressions for the invariants exist. There are indications
however that in order to achieve such a goal arrow diagrams as the ones
used in \cite{arrgpo} have to be introduced. The effect of the introduction
of these diagrams is to reduce the amount of embeddings entering the
product (\ref{producto}) to a selected set. Both, the expressions and the
amount of calculation could notably simplify if this is possible.
This issue is under investigation. 

Our approach opens a variety of investigations. First of all a
generalization of the reconstruction procedure from the kernels
(\ref{nucleos}) presented in \cite{temporal} up to order four should be
constructed. This could lead to a universal combinatorial formula for
Vassiliev invariants. The approach is also well suited to obtain
combinatorial expressions for Vassiliev invariants for links, a field which
has not been much explored up to now. Another context in which our approach
could be also very powerfull is in the study of vacuum expectation values
of graphs, quantities that plays an important role in recent developments in
the canonical approach to quantum gravity \cite{griego}. Vassiliev
invariants for graphas constitute a rather unexplored field which could
lead to new sets of important invariants.

\vskip2cm
\begin{center} {\bf Acknowledgements}
\end{center}

\vspace{4 mm}

We would like to thank L. Alvarez-Gaum\'e and  M. Alvarez for helpful
discussions on Vassiliev invariants and on gauge fixing. We also thank Simon
Willerton for bringing  Gauss diagrams to our attention and for sending us a
copy of his Ph. D. thesis. J.M.F.L. would like to thank the organizers of
the workshop on ``New Developments in Algebraic Topology" for their kind
invitation and their hospitality. This work was supported in part by DGICYT
under grant PB96-0960, and by the EU Commission under the TMR grant
FMAX-CT96-0012.

\begin{table}[hp]
\begin{center}
\begin{tabular}{|c||c|c|c|c|c|c||c|c|c|c|}\cline{1-5} \cline{7-11}
  Knot & $\nu_2$ & $\nu_3$ & $\nu_4^1$ & $\nu_4^2$  & $\;\;\;\;\;\;$
& Knot & $\nu_2$ & $\nu_3$ & $\nu_4^1$ & $\nu_4^2$ \\
  \cline{1-5} \cline{7-11}
 $3_1$ & 1 & 1 & 3 & 1 &  & $8_5$ & $-$1 & $-$3 & 1 & $-41$ \\
\cline{1-5} \cline{7-11}
 $4_1$ & $-$1 & 0 & 2 & -3 &  & $8_6$ & $-$2 & $-$3 & 7 & -36 \\
\cline{1-5} \cline{7-11}
 $5_1$ & 3 & 5 & 25 & 11  &  & $8_7$ & 2 & $-$2 & 4 & 22 \\
\cline{1-5} \cline{7-11}
 $5_2$ & 2 & 3 & 13 & 4 &  & $8_8$ & 2 & $-$1 & 5 & 12 \\
\cline{1-5} \cline{7-11}
 $6_1$ & $-$2 & $-$1 & 7 & $-12$ &  & $8_9$ & $-$2 & 0 & 14 & -34 \\
\cline{1-5} \cline{7-11}
 $6_2$ & $-$1 & $-$1 & 3 & $-13$ &  & $8_{10}$ & 3 & $-$3 & 15 & 15 \\
\cline{1-5} \cline{7-11}
 $6_3$ & 1 & 0 & 0 & 7 &  & $8_{11}$ & $-$1 & $-$2 & 2 & -27 \\
\cline{1-5} \cline{7-11}
 $7_1$ & 6 & 14 & 98 & 46 &  & $8_{12}$ & $-$3 & 0 & 14 & $-17$ \\
\cline{1-5} \cline{7-11}
 $7_2$ & 3 & 6 & 32 & 13 &  & $8_{13}$ & 1 & $-$1 & $-1$ & 17 \\
\cline{1-5} \cline{7-11}
 $7_3$ & 5 & 11 & 73 & 25 &  & $8_{14}$ & 0 & 0 & 4 & $-16$ \\
\cline{1-5} \cline{7-11}
 $7_4$ & 4 & 8 & 50 & 8 &  & $8_{15}$ & 4 & 7 & 37 & 18 \\
\cline{1-5} \cline{7-11}
 $7_5$ & 4 & 8 & 46 & 24 &  & $8_{16}$ & 1 & $-$1 & $-1$ & 17\\
\cline{1-5} \cline{7-11}
 $7_6$ & 1 & 2 & 8 & -1 &  & $8_{17}$ & $-$1 & 0 & 6 & $-19$ \\
\cline{1-5} \cline{7-11}
 $7_7$ & $-$1 & 1 & $-$1 & 3 &  & $8_{18}$ & 1 & 0 & 4 & $-9$ \\
\cline{1-5} \cline{7-11}
 $8_1$ & $-3$ & $-3$ & 13 &$-31$ & & $8_{19}$ & 5 & 10 & 60 & 35\\
\cline{1-5} \cline{7-11}
 $8_2$ & 0 & $-$1 & 3 & 30 &  & $8_{20}$ & 2 & 2 & 8 & 6 \\
\cline{1-5} \cline{7-11}
 $8_3$ & $-4$ & 0 & 30 & $-40$ &  & $8_{21}$ & 0 & $-$1 & $-$1 & $-14$ \\
\cline{1-5} \cline{7-11}
 $8_4$ & $-3$ & 1 & 21 & -39 &  &  &  &  &  &  \\
\cline{1-5} \cline{7-11}
\end{tabular}
\caption{Primitive Vassiliev invariants up to order four
for all prime knots up to eight crossings.}
\end{center}
\label{tablauno}
\end{table}

\begin{table}[hp]
\begin{center}
\begin{tabular}{|c||c|c|c|c|c|c||c|c|c|c|}\cline{1-5} \cline{7-11}
  Knot & $\nu_2$ & $\nu_3$ & $\nu_4^1$ & $\nu_4^2$ & $\;\;\;\;\;\;$
& Knot & $\nu_2$ & $\nu_3$ & $\nu_4^1$ & $\nu_4^2$  \\
  \cline{1-5} \cline{7-11}
 $9_1$ & 10 & 30 & 270 & 130 &  & $9_{26}$ & 0 & 1& $-$5 & 2 \\
\cline{1-5} \cline{7-11}
 $9_2$ & 4 & 10 & 62 & 32 &  & $9_{27}$ & 0 & 1 & 3 & $-6$  \\
\cline{1-5} \cline{7-11}
 $9_3$ & 9 & 26 & 228 & 87 &  & $9_{28}$ & 1 & 0 & $-$2 & 3 \\
\cline{1-5} \cline{7-11}
 $9_4$ & 7 & 19 & 151 & 51 &  & $9_{29}$ & 1 & $-$2 & 2 & 11 \\
\cline{1-5} \cline{7-11}
 $9_5$ & 6 & 15 & 115 & 20 &  & $9_{30}$ & $-$1 & $-$1 & 5 & $-9$ \\
\cline{1-5} \cline{7-11}
 $9_6$ & 7 & 18 & 134 & 77 &  & $9_{31}$ & 2 & 2 & 8 & 6 \\
\cline{1-5} \cline{7-11}
 $9_7$ & 5 & 12 & 78 & 47 &  & $9_{32}$ & $-$1 &2 & $-$2 & $-11$\\
\cline{1-5} \cline{7-11}
 $9_8$ & 0 & 2 & 8 & -8 &  & $9_{33}$ & 1 & $-$1 & 3 & 1 \\
\cline{1-5} \cline{7-11}
 $9_9$ & 8 & 22 & 180 & 80 &  & $9_{34}$ & $-$1 & 0 & 2 & $-3$ \\
\cline{1-5} \cline{7-11}
 $9_{10}$ & 8 & 22 & 188 & 48 &  & $9_{35}$ & 7 & 18 & 150 & 13 \\
\cline{1-5} \cline{7-11}
 $9_{11}$ & 4 & $-$9 & 57 & 10 &  & $9_{36}$ & 3 & $-$7 & 39 & 15 \\
\cline{1-5} \cline{7-11}
 $9_{12}$ & 1 & 3 & 15 & 1 &  & $9_{37}$ & $-$3 & 1 & 13 & $-7$ \\
\cline{1-5} \cline{7-11}
 $9_{13}$ & 7 & 18 & 142 & 45 &  & $9_{38}$ & 6 & 14 & 98 & 46 \\
\cline{1-5} \cline{7-11}
 $9_{14}$ & $-$1 & 2 & $-$6 & 5 &  & $9_{39}$ & 2 & $-$4 & 24 &  $-10$ \\
\cline{1-5} \cline{7-11}
 $9_{15}$ & 2 & $-$5 & 25 & 4 &  & $9_{40}$ & $-$1 & $-$1 & 3 & $-13$  \\
\cline{1-5} \cline{7-11}
 $9_{16}$ & 6 & 14 & 94 & 62 &  & $9_{41}$ & 0 & 1 & $-$9 & 18 \\
\cline{1-5} \cline{7-11}
 $9_{17}$ & $-$2 & 0 & 6 & $-2$ &  & $9_{42}$ & $-$2 & 0 & 10 & $-18$ \\
\cline{1-5} \cline{7-11}
 $9_{18}$ & 6 & 15 & 107 & 52 &  & $9_{43}$ & 1 & 2 & 14 & $-13$ \\
\cline{1-5} \cline{7-11}
 $9_{19}$ & $-$2 & 1 & 3 & 4 &  & $9_{44}$ & 0 & 1 & $-1$ & 10 \\
\cline{1-5} \cline{7-11}
 $9_{20}$ & 2 & 4 & 20 & 6 &  & $9_{45}$ & 2 & $-$4 & 20 & 6 \\
\cline{1-5} \cline{7-11}
 $9_{21}$ & 3 & $-$6 & 36 & $-3$ &  & $9_{46}$ & $-$2 & $-$3 & 3 & $-20$ \\
\cline{1-5} \cline{7-11}
 $9_{22}$ & $-$1 & 1 & 1 & 7 &  & $9_{47}$ & $-$1 & $-$2 & $-$6 & 5  \\
\cline{1-5} \cline{7-11}
 $9_{23}$ & 5 & 11 & 69 & 41 &  & $9_{48}$ & 3 & $-$5 & 29 & $-5$\\
\cline{1-5} \cline{7-11}
 $9_{24}$ & 1& 2 & 6 & $-5$  &  & $9_{49}$ & 6 & 14 & 102 & 30 \\
\cline{1-5} \cline{7-11}
 $9_{25}$ & 0 & 1 & 11 & $-14$  &  &  & & & & \\
\cline{1-5} \cline{7-11}
\end{tabular}
\caption{Primitive Vassiliev invariants up to order four 
for all prime  knots with nine crossings.}
\end{center}
\label{jpast}
\end{table}

\end{document}